\documentclass[pra,twocolumn,superscriptaddress]{revtex4}
\usepackage{graphicx,amsmath}

\begin{document}

\newcommand{\Lg}[0]{\ensuremath{{\cal L}_{\rm int}}}
\newcommand{\Lv}[0]{\ensuremath{{\cal V}}}
\newcommand{\Proj}[0]{\ensuremath{{\cal P}}}
\newcommand{\Ploc}[0]{\ensuremath{{\cal P}^{\rm loc}}}
\newcommand{\Lloc}[0]{\ensuremath{{\cal L}_{\rm loc}}}
\newcommand{\Lc}[0]{\ensuremath{{\cal L}}}
\newcommand{\Leff}[0]{\ensuremath{{\cal L}_{\rm eff}}}
\newcommand{\W}[0]{\ensuremath{{\cal W}}}
\newcommand{\T}[1]{\ensuremath{{\cal T}^{(#1)}}}
\newcommand{\Lpair}[0]{\ensuremath{{\cal L}_2^{\rm pair}}}

\newcommand{\ket}[1]{\ensuremath{\vert{#1}\rangle}}
\newcommand{\bra}[1]{\ensuremath{\langle{#1}\vert}}
\newcommand{\braket}[1]{\ensuremath{\langle{#1}\rangle}}

\newcommand{\kket}[1]{\ensuremath{\Vert{#1}\rangle\rangle}}
\newcommand{\bbra}[1]{\ensuremath{\langle\langle{#1}\Vert}}
\newcommand{\ketbra}[2]{\ensuremath{|#1\rangle\langle#2|}}
\newcommand{\bbrakket}[2]{\ensuremath{({#1}\vert{#2})}}

\renewcommand{\kket}[1]{\ensuremath{\vert{#1})}}
\renewcommand{\bbra}[1]{\ensuremath{({#1}\vert}}
\renewcommand{\ketbra}[2]{\ensuremath{|#1)(#2|}}

\title{Dissipation induced Tonks-Girardeau gas in an optical lattice}

\date{\today}

\author{J.~J. \surname{Garc\'{\i}a-Ripoll}}
\affiliation{Facultad de Ciencias F\'{\i}sicas,
  Universidad Complutense de Madrid,
  Ciudad Universitaria, E-28040, Madrid, Spain}
\affiliation{Max-Planck-Institut f\"{u}r Quantenoptik,
  Hans-Kopfermann-Str.\ 1, 85748 Garching, Germany}
\author{S. \surname{D\"urr}}
\affiliation{Max-Planck-Institut f\"{u}r Quantenoptik,
  Hans-Kopfermann-Str.\ 1, 85748 Garching, Germany}
\author{N. \surname{Syassen}}
\affiliation{Max-Planck-Institut f\"{u}r Quantenoptik,
  Hans-Kopfermann-Str.\ 1, 85748 Garching, Germany}
\author{D.~M. \surname{Bauer}}
\affiliation{Max-Planck-Institut f\"{u}r Quantenoptik,
  Hans-Kopfermann-Str.\ 1, 85748 Garching, Germany}
\author{M. \surname{Lettner}}
\affiliation{Max-Planck-Institut f\"{u}r Quantenoptik,
  Hans-Kopfermann-Str.\ 1, 85748 Garching, Germany}
\author{G. \surname{Rempe}}
\affiliation{Max-Planck-Institut f\"{u}r Quantenoptik,
  Hans-Kopfermann-Str.\ 1, 85748 Garching, Germany}
\author{J.~I. \surname{Cirac}}
\affiliation{Max-Planck-Institut f\"{u}r Quantenoptik,
  Hans-Kopfermann-Str.\ 1, 85748 Garching, Germany}

\begin{abstract}
  We present a theoretical investigation of a lattice Tonks-gas that
  is created by inelastic, instead of elastic interactions. An
  analytical calculation shows that in the limit of strong two-body
  losses, the dynamics of the system is effectively that of a Tonks
  gas. We also derive an analytic expression for the effective loss
  rate. We find good agreement between these analytical results and
  results from a rigorous numerical calculation. The Tonks character
  of the gas is visible both in a reduced effective loss rate and in
  the momentum distribution of the gas.
\end{abstract}

\maketitle

\section{Introduction}

A Tonks-Girardeau gas is a one dimensional (1D) gas of bosons where a
repulsive interaction dominates all other energy scales. It was
predicted long ago \cite{girardeau:60,lieb:63} that repulsion can
mimic the Pauli exclusion principle and that, for one dimension, the
wavefunction for bosons and fermions would be the same, except for a
simple transformation. Not only that, but the excitation spectrum of
these one-dimensional hard core bosons can also be mapped to that of a
fermionic system. A few years ago, experiments observed a
Tonks-Girardeau gas using ultracold gases in optical lattices with
strong repulsive interactions
\cite{paredes:04,kinoshita:04,kinoshita:05}.

In a recent experiment \cite{syassen:08} we showed that strong
dissipation in the form of two-body losses can also simulate a Pauli
exclusion principle, fermionizing a system and transforming it into a
dissipative but long-lived strongly correlated gas. This equivalence
was demonstrated for molecules of $^{87}$Rb loaded in an optical
lattice. For deep enough lattices, these particles exhibit strong
two-body decay rates which, in appropriate units, exceed the average
kinetic energy of the particles. The particles then avoid coming
close together and behave like impenetrable bosons. This happens both in the
continuum case of 1D tubes and when the tubes are modulated by a deep
optical lattice ---two situations which resemble the experiments with
elastically interacting bosonic atoms in
Refs.~\cite{kinoshita:04,kinoshita:05} and Ref.~\cite{paredes:04},
respectively.

\begin{figure}
  \centering
  \includegraphics[width=\linewidth]{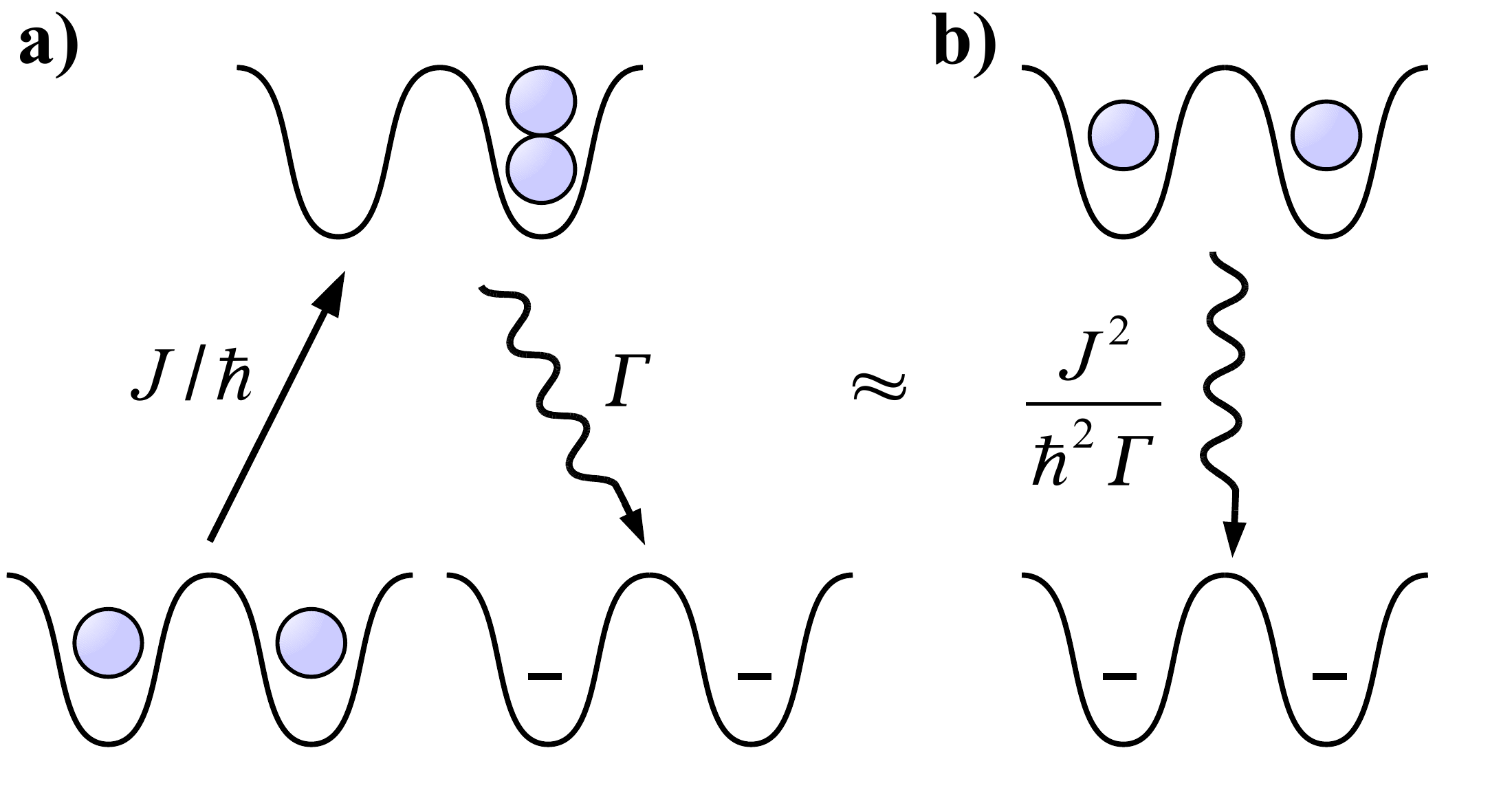}
  \caption{Two molecules sit on neighboring lattice sites. The
    molecules can tunnel with a hopping amplitude $J/\hbar$ which is
    much weaker than the decay rate of the resulting doubly occupied site
    into the vacuum. The effective model is that of impenetrable
    particles decaying with a much weaker ${\cal
      O}(J^2/\hbar^2\Gamma)$ decay rate.}
  \label{fig:zeno}
\end{figure}

At least in the lattice case, the equivalence between strong
dissipation and a Pauli exclusion principle can be understood in terms
of the Zeno effect. Following the discussion in
Ref.~\cite{syassen:08}, Fig.~\ref{fig:zeno} depicts a stable
configuration with two molecules on neighboring sites. This
configuration is connected via hopping to states with double
occupancy. These states decay at a rate $\Gamma$ which is much larger
than the hopping amplitude $J/\hbar$ of the particles. Treating this
as a typical three-level system from Quantum Optics, one concludes
that particles stay on their original sites with only a minor loss
rate of ${\cal O}(J^2/\hbar^2\Gamma).$

In this work we present a rigorous theoretical analysis of this
system. The outline is as follows. We begin in
Section~\ref{sec:formalism} by introducing a master equation that
models two-body losses for a single species of bosonic particles --for
instance the molecules in Ref.~\cite{syassen:08}--. We will
particularize the model to the case in which particles are confined by
an optical lattice and explain how dissipation becomes the dominant
term. In Section~\ref{sec:effective} we will show that in the limit of
strong losses the master equation can be replaced by an effective
model in which the rapid two-body decay has been eliminated. The
dominant terms of the effective model are identical to the Hamiltonian
of an elastic Tonks-Girardeau gas. The residual effect of losses is a
slow perturbation that can be obtained analytically. In
Section~\ref{sec:numerics} we compare both results to exact numerical
simulations of the full master equation. We use Matrix Product Density
Operators \cite{verstraete:04} (MPDO) to verify that both the density
and momentum distributions of the particles closely resemble those of
a Tonks gas. A second signature of the hard-core boson dynamics is a
slowdown of the inelastic losses, for which we find good agreement
between numerical and analytical estimates. In the last part of our
work we offer further details about the methods and derivation used to
obtain the effective Tonks gas models and the slowdown of
losses. Finally, we summarize this paper in
Section~\ref{sec:conclusions}.

\section{Formalism}
\label{sec:formalism}

\subsection{Master equation}

We model the dynamics of the particles using a Markovian master equation
\begin{eqnarray}
  \hbar \frac{d\rho}{dt} = - i [ H, \rho] + {\cal D}\rho ,
\end{eqnarray}
where $H$ is the Hamiltonian describing the unitary part of the
evolution and $\cal D$ is a dissipator associated with the losses due
to inelastic collisions. In the absence of a lattice potential
\cite{duerr:prep}
\begin{eqnarray}
  H &=& \int d^3x \Psi^\dagger H_s \Psi +
  \frac{{\rm Re}(g_{3D})}{2} \int d^3x \Psi^{\dagger2} \Psi^2 \label{eq:master}\\
  {\cal D} \rho &=& - \frac{{\rm Im}(g_{\rm 3D})}{2} \nonumber \\
  && \times \int d^3x \left( 2 \Psi^2 \rho \Psi^{\dagger 2} - \Psi^{\dagger 2} \Psi^2 \rho 
    - \rho \Psi^{\dagger 2} \Psi^2 \right) ,
\end{eqnarray}
where $H_s({\bf x})=-\hbar^2\nabla^2/2m+V_{\rm trap}({\bf x})$ is the
single-particle Hamiltonian and $\Psi({\bf x})$ is the bosonic field
operator. The strength of the interparticle interactions is
$g_{3D}=4\pi\hbar^2 a/m$, where $m$ is the mass of a particle and $a$
is the scattering length. ${\rm Re}(a)$ describes elastic collisions,
whereas ${\rm Im}(a)\leq 0$ describes inelastic collisions that lead
to losses.

To assert the consistency of this model, let us estimate the decay
rate of the particles. When taking expectation values over the density
operator, $n(x)=\Psi^\dagger({\bf x})\Psi({\bf x}),$ the previous
master equation becomes
\begin{equation}
  \frac{d}{dt} \braket{\Psi^\dagger({\bf x})\Psi({\bf x})} =
  \frac{2}{\hbar}\, {\rm Im}(g_{\rm 3D})
  \braket{\Psi^{\dagger 2}({\bf x})\Psi^2({\bf x})}.\label{eq:decay}
\end{equation}
For a condensate of molecules this gets the expected form
\begin{equation}
  \frac{d}{dt} \bar n({\bf x}) \simeq -K_{\rm 3D}  \bar n({\bf x})^2,
\end{equation}
where $\bar n({\bf x})$ is the density of bosons at a given point and
the decay rate is proportional to the square of the density and to
$K_{\rm 3D} = -2\,\mathrm{Im}(g_{3D})/\hbar.$

\subsection{Optical lattice}

As explained in the introduction, in this paper we want to model the
experiment with molecules in a deep optical lattice
\cite{syassen:08}. For low temperatures and tight confinements, we can
expect that the particles will accommodate to the motional ground
state of each lattice site. Note that if this is true for individual
atoms, it is even more so for molecules, because having twice the mass
and twice the polarizability, they are better trapped by the same
optical potential.  Under these conditions, it was shown in
Ref.~\cite{jaksch:98} that it is convenient to expand the bosonic
field operator $\Psi^\dagger$ using Fock operators, $a^\dagger_k,$
that create particles on the $k$-th lattice site:
\begin{equation}
  \Psi^\dagger({\bf x}) = \sum_k a_k^\dagger w({\bf x}-{\bf x}_k).
\end{equation}
The $w({\bf x})$ are Wannier wavefunctions associated to the states
localized on each site. In particular, for the arrays of
one-dimensional lattices used in Ref.~\cite{syassen:08}, we can
separate the wavefunction of these bosonic modes into a product of
three wavepackets, $w(\mathbf{x}) = w_{\Vert}(x)
w_\perp(y)w_\perp(z),$ a less confined longitudinal one, $w_\Vert,$
and two transverse ones, $w_\perp,$ which are very tight due to the
perfect decoupling between adjacent tubes.

The advantage of the Wannier functions is that we can now perform a
tight-binding approximation: anywhere outside the single-particle
terms in the Hamiltonian, the overlap of different Wannier functions
is neglected.  This procedure transforms the unitary part of the
master equation into a Bose-Hubbard model \cite{jaksch:98}, and
discretizes the dissipative terms as well
\begin{eqnarray}
  H &=& -J\sum_{\braket{k,l}} a^\dagger_k  a_l+
  \frac {U_r} 2  \sum_k a_k^{\dagger 2}a_k^2, \label{H-tunnel}\\
  {\cal D} \rho &=&  \frac{\hbar\Gamma}4 \sum_k(2 a^2_k\rho a_k^{\dagger 2} -
  a^{\dagger 2}_ka^2_k\rho - \rho a^{\dagger 2}_ka_k^2).\nonumber
\end{eqnarray}
Regarding the notation, the sum $\langle k,l \rangle$ extends over
nearest neighbors along the same tube, $|k-l|=1.$ The tunneling
amplitude between neighboring lattice sites is denoted by $J$. Finally,
the on-site interaction matrix element contains both a real and
imaginary part
\begin{eqnarray}
  U &=& g_{3D} \int dx |w_{\Vert}(x)|^4 \left[\int dy|w_\perp(y)|^4\right]^2\\
   &=& U_r + i U_i = U_r - i \frac{\hbar \Gamma}{2},\nonumber
\end{eqnarray}
which contribute to the unitary and the dissipative parts of the
master equation, respectively.

The imaginary part of the interaction constant governs the decay of
the number of particles per site, $n_k=a^{\dagger}_ka_k$
\begin{equation}
  \frac{d}{dt}\braket{n_k} = - \Gamma \braket{n_k(n_k-1)}.
\end{equation}
In the cases that we will study this rate will be larger than the
speed at which particles tunnel to neighboring sites, $\Gamma \gg
J/\hbar.$ To facilitate the calculations, we will group the terms in
the master equation according to their strength. We introduce a
superoperator $\Lv$ that contains the tunneling part ($H_J\propto J$)
and a superoperator $\Lg$ that describes the elastic ($H_{\rm
  el}\propto U_r$) and inelastic interactions (${\cal D} \propto
\Gamma$)
\begin{subequations}
  \label{master-eq}
\begin{eqnarray}
  \label{master-eq-V-Lint}
  \frac{d}{dt}\rho &=& \left(\Lv + \Lg \right)\rho \\
  \label{Lv-def}
  \Lv \rho &=& -\frac{i}{\hbar} \left[H_J, \rho\right] \\
  \Lg \rho &=& -\frac{i}{\hbar} \left[H_{\rm el},\rho\right] +
  \frac1\hbar {\cal D} \rho .
\end{eqnarray}
\end{subequations}
The term ``superoperator'' refers to the fact that $\cal D$, $\Lv$,
and $\Lg$ are linear operators acting on density matrices, not on pure
states. Finally, in the following it will be important to realize that
$\Lv$ is of order $J$, whereas $\Lg$ is of order $|U| \gg |J|,$ and
dominates the evolution. The following sections exploit this
difference of scales to create new and simpler effective models that
describe the dynamics of the molecules.

\section{Effective models}
\label{sec:effective}

\subsection{Second order effective theory}

Our goal is to develop an effective master equation that is equivalent
to (\ref{master-eq}) in the limit of strong dissipation, $\hbar\Gamma
\gg J,$ in which $\Lg$ dominates. We will sketch the main ideas
and refer the reader to Sect.~\ref{sec:calculations} for more
details. The process begins by identifying the eigenvalues,
$\lambda_i,$ and eigenspaces of the dominant term,
$\Lg=\sum_i\lambda_i\Proj_i.$ We then decompose the density matrix
into a sum of contributions from these eigenspaces
\begin{equation}
  \rho(t) = \sum_i \rho_i(t) = \sum_i \Proj_i\rho(t).
\end{equation}
Finally, we will determine some approximate evolution equations for
the different $\rho_i$ in the presence of a non-zero hopping term,
$\Lv.$

We only need to study three eigenspaces, corresponding to the
dissipation-less states, $\lambda_0=0,$ and to the states most
immediately connected to them by the hopping. The most relevant term
of our density matrix, $\rho_0 \sim {\cal O}(1),$ is given by states
with zero or one particle per site. These states do not decay in the
absence of hopping and have the property that they belong to a space
with zero or one particle per site. We thus have
\begin{equation}
  \rho_0 = Q_0\rho Q_0, \label{stable}
\end{equation}
with a projector expressed in the Fock basis of occupation numbers
\begin{equation}
  Q_0 = \left( \ket{0}\bra{0} + \ket{1}\bra{1} \right)^{\otimes L}
  = q_0^{\otimes L},
\end{equation}
where $L$ is the number of sites in the lattice. The next states that
we need to consider have a pair of particles on some site. As we will
see later, we have two sets of states depending whether the double
occupation is on the left or on the right side of the matrix
coherences
\begin{eqnarray}
  \Proj_{1a}\rho &=& Q_1 \, \rho \, Q_0,\\
  \Proj_{1b}\rho &=& Q_0 \, \rho \, Q_1,\nonumber
\end{eqnarray}
where we have introduced
\begin{equation}
  Q_1 = \sum_{k=0}^L
  q_0^{\otimes k-1} \otimes \ket{2}\bra{2} \otimes q_0^{L-k}.
\end{equation}
In the absence of hopping these states decay with eigenvalues
$\lambda_{1a}=-\Gamma/2-i U_r/\hbar=-iU/\hbar$ and $\lambda_{1b}=-\Gamma/2+i
U_r/\hbar = i U^{*}/\hbar,$ respectively

By neglecting higher order contributions to the density matrix, it is
possible to integrate formally the projected master equation and
obtain an effective model for the dominant term, $\rho_0(t).$ After
some manipulations one arrives to the following model
[Sect.~\ref{sec:calc-effective}]
\begin{subequations}
  \label{effective-model}
\begin{eqnarray}
  \frac{d\rho_0}{dt} &=& (\Lc_1 + \Lc_2) \rho_0 \\
  \Lc_1 &=& \Proj_0\Lv\Proj_0  \\
  \Lc_2 &=& \sum_{c\in\{1a,1b\}}
  \frac{-1}{\lambda_c}\Proj_0\Lv\Proj_c\Lv\Proj_0,
  \label{L2}
\end{eqnarray}
\end{subequations}
We will now write down explicitly and discuss the meaning and
implications of these different terms.

\subsection{Tonks-Girardeau Gas}
\label{sec:tonks}

The most important term in our effective model (\ref{effective-model})
is given by the Liouville operator $\Lc_1=\Proj_0\Lv\Proj_0.$ As shown in
Sect.~\ref{sec:calc-tonks} this superoperator is equivalent to a
Hamiltonian of hard-core bosons, also known as a Tonks-Girardeau gas.
Therefore, to lowest order
\begin{eqnarray}
  \label{Tonks}
  \frac{d}{dt}\rho_0 = -\frac i\hbar \left[-J\sum_{\braket{k,l}}c^{\dagger}_kc_l,\rho_0\right]
  +  {\cal O}\left(J^2/|U|\right),
\end{eqnarray}
with bosonic hard-core operators $c^\dagger_k$ and $c_k$ that have
implicitely the restriction of one particle per site
\begin{equation}
  c_k = \ket{0}_k\bra{1}_k,\quad c_k^\dagger = \ket{1}_k\bra{0}_k.
\end{equation}

This is the main result of our paper. Namely, that a strong
dissipation such as the two-body losses from our system can lead to
coherent evolution. As it was mentioned in the introduction, the same
result can be obtained in an alternative way. By establishing an
analogy between losses and a continuous measurement, it is intuitively
clear that the strong dissipation causes a Zeno effect which
suppresses the process of two particles coming together and being
lost. In this regime, the dynamics of the molecules must be given by a
Hubbard model where doubly occupied states have been projected out.

In practice, the hard-core bosons model implies a very simple dynamics
that can be tested numerically, as we do in
Section~\ref{sec:numerics}. However, verifying the same thing with
real particles in an optical lattice represents a challenging
experiment. Nevertheless, one can easily check two phenomena: first,
that in the regime of strong decay, $\hbar\Gamma \gg J,$ no
significant losses take place in a time scale of order $1/\Gamma;$ and
second, that the effective loss rate can be estimated using the second
part of our effective model (\ref{effective-model}c). This is the goal
of the following sections.

\subsection{Second order corrections}
\label{sec:losses}

With a lengthy calculation we can rewrite the second order terms in
Eq.~(\ref{effective-model}) as an effective Liouvillian $\Lc_2$ with
both a Hamiltonian and dissipation
\begin{equation}
  \Lc_2 \rho_0 = -\frac{i}{\hbar}[H_2,\rho_0] + \frac 1{\hbar}
  {\cal D}_2\rho_0.
\end{equation}
It is convenient to introduce an operator that destroys a pair of particles
in neighboring sites
\begin{equation}
  C_k = c_k(c_{k+1}+ c_{k-1}). \label{pairs}
\end{equation}
With these pairs the Hamiltonian part can be written
\begin{equation}
  H_2 = -J_2 \sum_k C^\dagger_k C_k,
\end{equation}
with an effective strength $(1/\lambda_{1a}= \hbar i/U)$
\begin{equation}
  J_2 = \frac{2J^2}{\hbar}\mathrm{Im}\left(\frac{1}{\lambda_{1a}}\right)
    = \frac{2J^2}{|U|^2} U_r.
\end{equation}
Note that this Hamiltonian contains both effective nearest neighbor
interactions and a three-site hopping of the form
$c^\dagger_{k+1}n_kc_{k-1}.$ Both terms are typical of the
Bose-Hubbard model in the limit of strong repulsive interaction $U_r
\gg |J|$ \cite{freericks:96}. Moreover, these Hamiltonian corrections
disappear when the lossy particles do not interact elastically
on-site, $U_r=0.$

The dissipative term is equally simple,
\begin{equation}
  {\cal D}_2\rho_0 = \hbar \Gamma_2 \sum_k
  \left(2C_k\rho_0C_k^\dagger-C^\dagger_kC_k\rho_0 - \rho_0C^\dagger_kC_k
  \right),
\end{equation}
and has a loss coefficient
\begin{equation}
  \Gamma_2 = -\frac{2J^2}{\hbar^2}\mathrm{Re}\left(\frac{1}{\lambda_{1a}}
    \right) = -\frac{2J^2}{\hbar|U|^2}U_i.
\end{equation}
In the limit of weak elastic interaction between molecules we may
write
\begin{equation}
  \Gamma_2 =  -\frac{2J^2}{\hbar U_i}\left(1+\frac{U_r^2}{U_i^2}\right)^{-1}
  = \frac{4J^2}{\hbar^2 \Gamma}
  \left(1+\frac{4U_r^2}{\hbar^2\Gamma^2}\right)^{-1},
\end{equation}
which shows that the decay rate is indeed ${\cal O}(J^2/\hbar^2\Gamma)$
as anticipated.

\subsection{Effective losses}

Let us write the evolution of the total number of particles, $\hat N,$
under the effective master equation (\ref{effective-model})
\begin{equation}
  \frac{d}{dt}\braket{\hat N} = -\sum_k \Gamma_2
  \braket{[C^\dagger_kC_k,\hat N]}
  = -4\Gamma_2 \sum_k \braket{C^\dagger_kC_k},
\end{equation}
This expression in general cannot be simplified any further, at least
not without some assumption about the state with which we compute the
expectation value.

The experiments described in Ref.~\cite{syassen:08} start from a state
with exactly one molecule at each site and evolve only until
approximately half of the particles are lost. Here further
approximations are possible: first we treat the system as homogeneous
and second we assume that the populations of different sites are
uncorrelated{\footnote{Note that this does not imply that the
    particles themselves are uncorrelated}}. We thus obtain
\begin{equation}
  \braket{C^\dagger_kC_k} =
  \sum_{l,l'\in\{k-1,k+1\}}
  \braket{c^\dagger_kc^\dagger_{l}c_kc_{l'}} \simeq z\bar n^2,
\end{equation}
where $\bar n$ is the density and $z=2$ is the coordination number of
our lattice.  This approximation leads to a rate equation
characteristic of two-body processes
\begin{equation}
  \frac{d}{dt}\bar n \simeq
  -4 z \Gamma_2 \bar n^2 \equiv -\kappa
  \bar n^2.\label{two-body}
\end{equation}
A similar equation was derived previously in Ref.~\cite{syassen:08}
using a more restricted theory than our effective master equation
(\ref{effective-model}).

\section{Numerical Results}
\label{sec:numerics}

In order to study the quality of the approximations used in the
effective model, we have performed numerical simulations of the full
master equation (\ref{master-eq}) using MPDO
\cite{verstraete:04}. This is a method that approximates the density
matrix $\rho(t)$ using a matrix product state structure. As described
in Ref.~\cite{verstraete:04}, this variational ansatz is well suited
to simulating evolution of the state under a Liouvillian like $\Lv +
\Lg,$ which can be decomposed into a sum of local or nearest-neighbor
terms. In our simulations we have worked with up to 48 sites and open
boundary conditions, setting $U_r=0$ so that the different effects
cannot be attributed to a repulsive interaction. We have experimented
with different cutoffs, from two to four particles per site, verifying
that they give similar results.

\begin{figure}
\includegraphics[width=\linewidth]{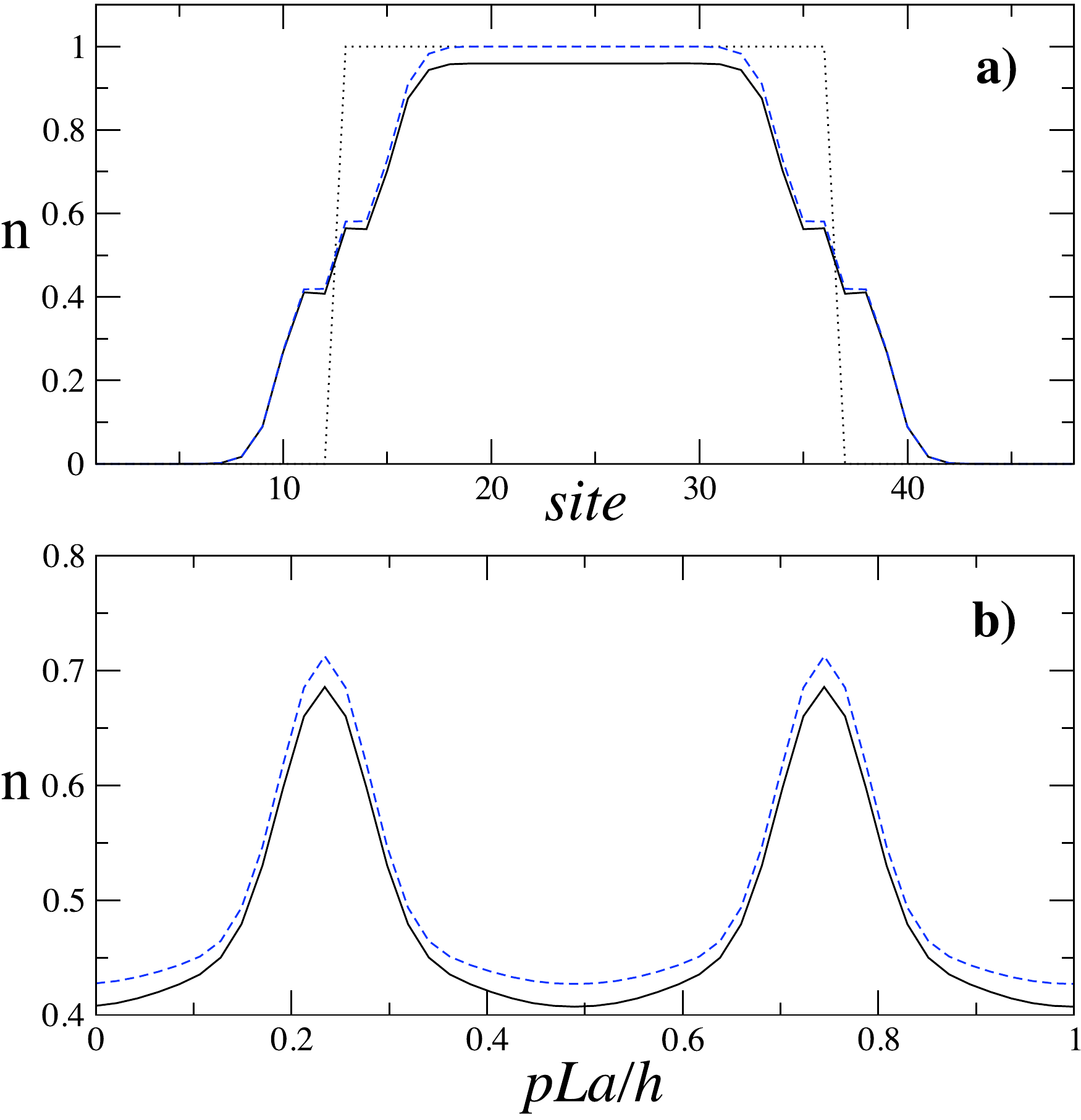}
\caption{\label{fig:profile} (Color online) Comparison between the
  dissipative system (solid lines) and the lossless Tonks gas (dashed
  lines). $N=24$ central sites of the lattice initially contain
  exactly one particle per site [dotted line in (a)]. The system is
  evolved with $\hbar\Gamma/J = 4000$ for a time $t = 2\hbar/J$. The
  evolution is simulated either with the full master equation
  (\ref{master-eq}) or with the lossless Tonks model Eq.\
  (\ref{Tonks}). Part (a) shows the position distribution versus site
  index, part (b) the momentum distribution versus quasi-momentum $p$
  ($a$ is the lattice spacing and $L$ the lattice length), both after
  the time evolution. The difference between the distributions is
  marginal, except for an overall reduction of particle number.  }
\end{figure}

\subsection{Comparison with the Lossless Tonks Gas}

Fig.~\ref{fig:profile} shows numerical results for $\hbar
\Gamma/J=4000$. The system was initially prepared such that the $N=24$
central sites of the lattice contain one particle per lattice site,
while all other lattice sites are empty, similar to the state
experimentally prepared in Ref.~\cite{volz:06}. We then let the system
evolve for a time $t=2\hbar/J$ and measure both the density and the
momentum distributions, which are plotted in
Figs.~\ref{fig:profile}(a) and \ref{fig:profile}(b), respectively. The
solid lines were obtained with the full master equation
(\ref{master-eq}) and differ only slightly from the dashed lines
obtained with the lossless Tonks model of Eq.~(\ref{Tonks}). The
observed difference in the distributions is largely due to the reduced
particle number.

\begin{figure}[tb!]
\includegraphics[width=\linewidth]{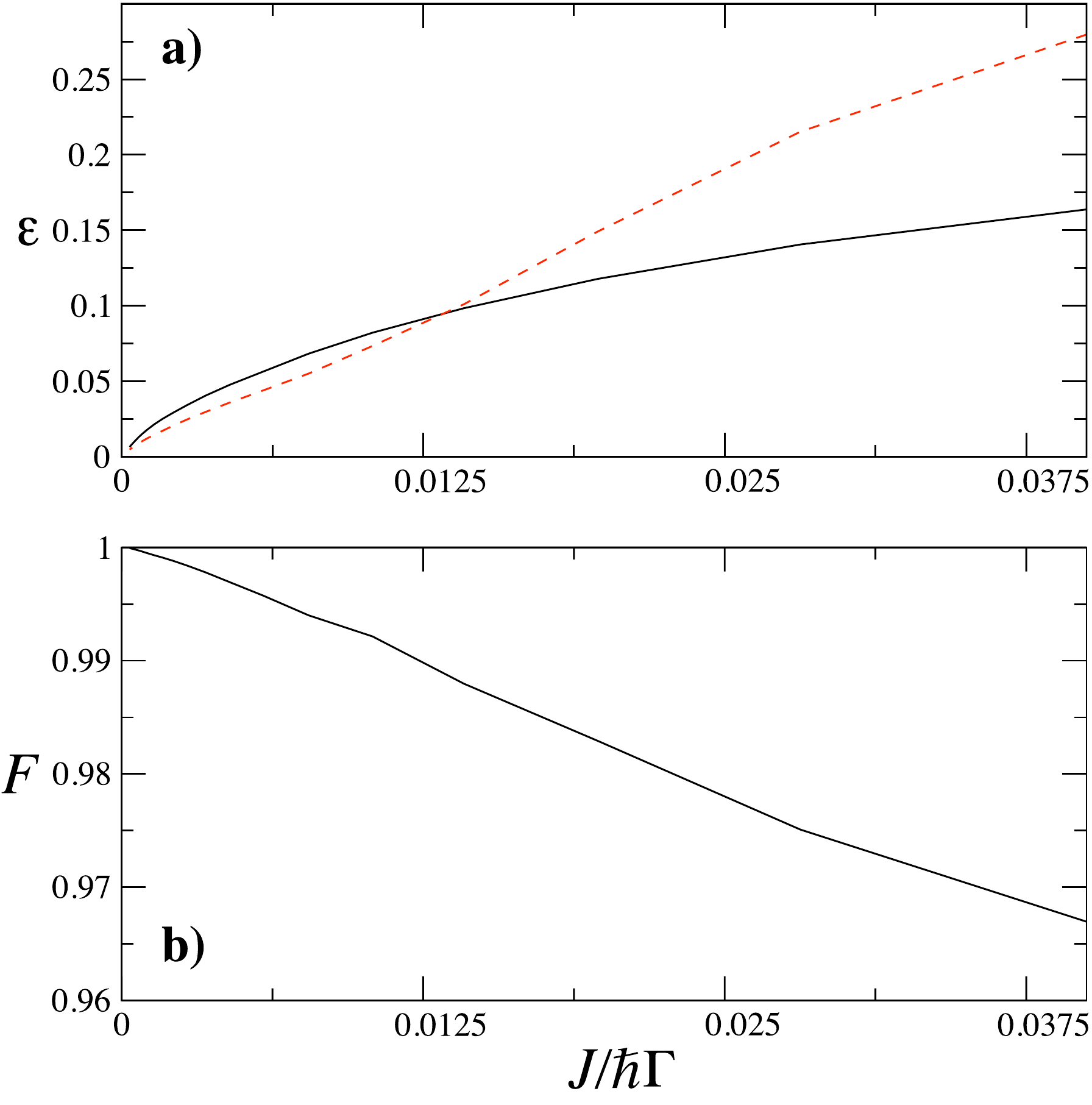}
\caption{\label{fig:fidelity} (Color online) (a) Difference between
  the distributions of the dissipative system and the lossless Tonks
  gas. Calculations as in Fig.~\ref{fig:profile} for a variety of
  values $J/\hbar\Gamma.$ Each yields a different distribution for the
  dissipative system and the lossless Tonks gas. The value $\epsilon$
  defined in Eq.\ (\ref{epsilon}) quantifies this difference for the
  position distribution (solid line) and the momentum distribution
  (dashed line). (b) Fraction of the density matrix in the hard-core
  bosons subspace, with $F$ defined in Eq. (\ref{fidelity}). $1-F$ is
  typically smaller than $\epsilon$.}
\end{figure}

In order to quantify how small the remaining difference of the
position distributions is, we define
\begin{eqnarray}
\label{epsilon}
  \epsilon = \sum_{x=1}^L\left|\frac{n_x}{N(t)} - \frac{n_x^{\mathrm{Tonks}}}{N(0)}\right|,
\end{eqnarray}
where $N(t)$ is the total number of particles at a given time, $n_x$
is the number of particles for the site $x$ and $n_x^{\mathrm{Tonks}}$
is the same for the lossless Tonks gas. A similar measure can be
defined in momentum space. Fig.~\ref{fig:fidelity}(a) shows these
quantities. For $J/\Gamma\rightarrow 0$ the distributions converge to
those of a lossless Tonks gas, as expected from the effective model.

Another important quantity in this comparison is the fraction of the
state in the hard-core bosons subspace of states with at most single
occupation
\begin{eqnarray}
  \label{fidelity}
  F = \mathrm{Tr}(Q_0\rho)
\end{eqnarray}
This quantity is shown in Fig.~\ref{fig:fidelity}(b). Comparison with
part (a) shows that $1-F$ tends to be much smaller than
$\epsilon$. This is because loss causes the system to evolve into an
incoherent mixture of systems with different total particle
number. Each of these systems remains to a pretty good approximation
in the Tonks gas subspace, but the position and momentum distributions
look different so that the mere scaling with total particle number
yields larger deviations in $\epsilon$.

\subsection{Effective Loss Rate}

\begin{figure}[tb!]
\includegraphics[width=\linewidth]{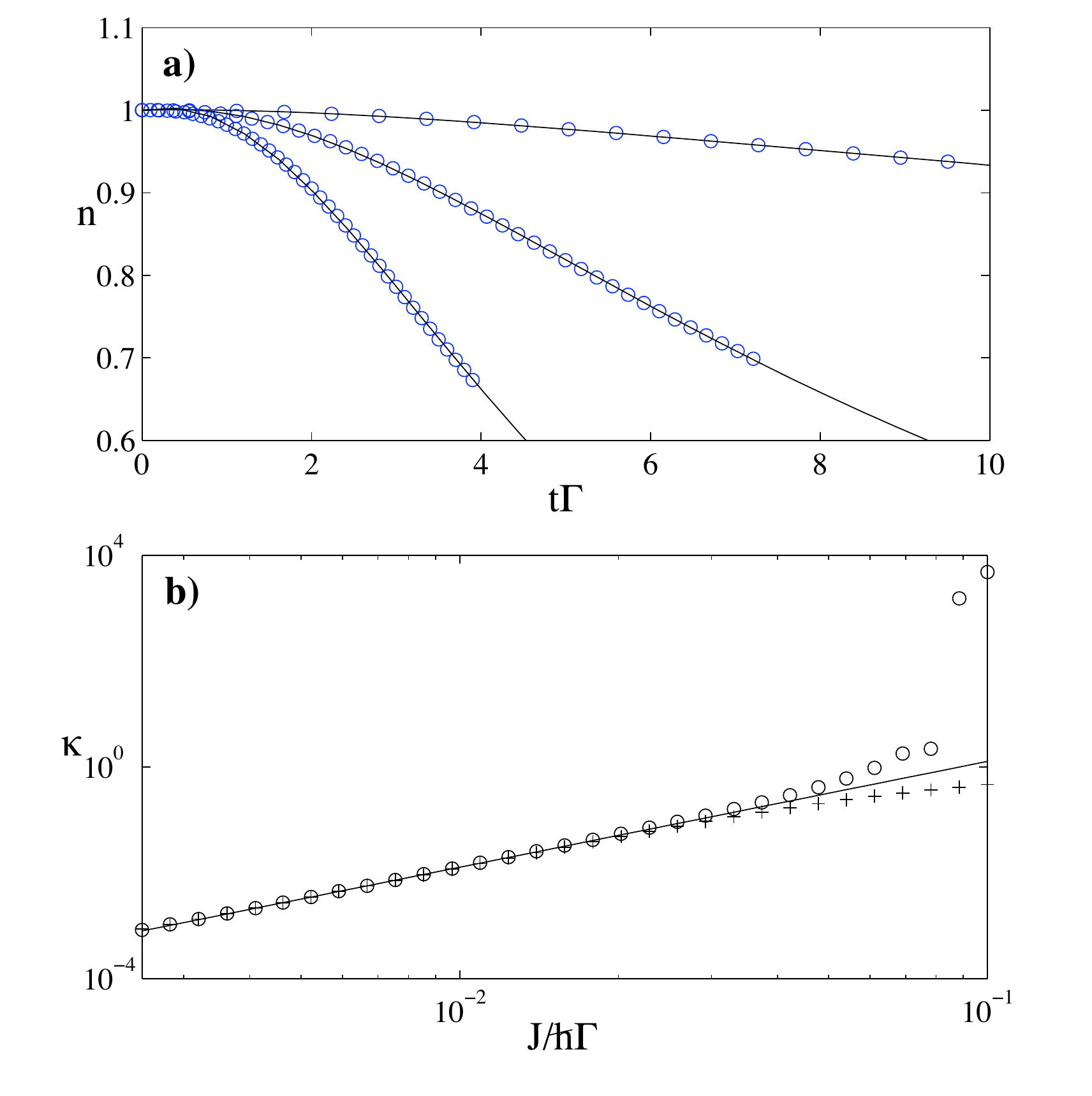}
\caption{\label{fig:two-body-decay} (Color online) We study the
  density as a function of time, for a simulation of
  Eq.~(\ref{master-eq}) using $L=40$ sites, each one initially filled
  with exactly one particle. (a) Evolution of the density as a
  function of the adimensionalized time $t\Gamma,$ for three values of
  the losses $J/\hbar\Gamma=0.0179$, $0.0541$ and $0.1$ (from top to
  bottom). We present data from both the numerical simulation and for
  a fit with Eq.~(\ref{fit1}), using circles and lines,
  respectively. (b) Best-fit values of the effective decay rate
  $\kappa$ vs. the value of $J/\hbar\Gamma$ at which the simulation
  took place. The solid line represents the theoretical prediction
  Eq.~(\ref{two-body}). The crosses and the circles corresponds to
  fits using the formulas in Eqs.~(\ref{fit2}) and (\ref{fit1}),
  respectively.}
\end{figure}

Our analytic model suggests that the loss rate of particle number in a
homogeneous system can be approximated by a two-body decay
Eq.~(\ref{two-body}), which corresponds to an evolution of the density
given by
\begin{equation}
  \bar n(t) = \frac{\bar n(0)}{1 + \bar n(0) \kappa t},
  \label{fit}
\end{equation}
This result applies to the case in which different sites are not very
much correlated and the losses $\Gamma$ dominate over the hopping,
$J.$

In order to test this prediction, we have performed MPDO simulation of
Eq.~(\ref{master-eq}) using a uniformly filled lattice with $L=40$
sites and $N=40$ particles. We studied the evolution for a wide range
of values of $J/\hbar\Gamma,$ with some examples shown in
Fig.~\ref{fig:two-body-decay}(a). After an initial transient that
vanishes on a timescale $\approx 1/\Gamma$, a two-body decay
Eq.~(\ref{two-body}) fits the data fairly well. However, while for
$J/\hbar\Gamma\ll 1$ the effect of the transient is negligible, for
larger $J/\hbar\Gamma$ there are better ways to fit the resulting
curves. One possibility, used in Ref.~\cite{syassen:08} to fit the
numerical data, is to include a free parameter $t_0$ describing an
offset on the time axis
\begin{equation}
  \label{fit2}
  \bar n(t) = \frac{\bar n(0)}{1 + \bar n(0) \kappa (t - t_0)}.
\end{equation}
However, we found that a more accurate model is a modified decay
equation with an exponentially modulated decay coefficient
\begin{equation}
  \frac{d}{dt}\bar n = -\kappa (1 - e^{-\lambda t}) \bar n^2.
\end{equation}
Here the exponential term with $\lambda \sim \Gamma$ represents an
heuristic approximation to the transients that we have neglected in
developing the effective model [See Sect.~\ref{sec:calc-effective}].
The solution of this differential equation still has two fit
parameters
\begin{equation}
  \bar n(t) = \frac{\bar n(0)}
  {1 + \bar n(0) \kappa \{t + [\exp(-\lambda t) - 1]/\lambda\}}.
  \label{fit1}
\end{equation}
Fig.~\ref{fig:two-body-decay}(b) shows the loss rate coefficient
$\kappa$ from these two fits and Fig.~\ref{fig:two-body-decay}(a)
shows the quality of the fits for small values of the hopping. For
larger values, though, the fitting becomes numerically unstable. In
this regime a fit (\ref{fit2}) behaves better.

What we do not show in the previous plots is that the long term
behavior of the system no longer follows the simple two-body decay
laws from Eqs.~(\ref{fit2}) and (\ref{fit1}). The reason is that at
low densities there are enough correlations that we can no longer use
the simple models from Section~\ref{sec:losses}. In this regime ofhe
small densities, a coarse grained description becomes approximately
equivalent to the Lieb-Liniger model \cite{cazalilla03} but with
inelastic interactions. As we have shown elsewhere \cite{syassen:08},
the decay at long times is then expected to follow the law $d \bar
n/dt \propto \bar n^4.$

\section{Detailed calculations}
\label{sec:calculations}

The goal of this work is to find an effective model for the particles
in the lattice, which works in the limit of fast dissipation, $J \ll
\hbar\Gamma.$ Our main tool to understand this limit is a
generalization of Kato perturbation theory \cite{kato:80}, also known
as adiabatic elimination, to the superoperators $\Lv$ and $\Lg.$
Section~\ref{sec:strategy} explains how our calculations relate to
this broader scope. Readers less interested in this aspect may skip to
Section~\ref{sec:projectors} where the actual derivation begins.

\subsection{General ideas}
\label{sec:strategy}

\begin{figure}
\includegraphics[width=0.8\linewidth]{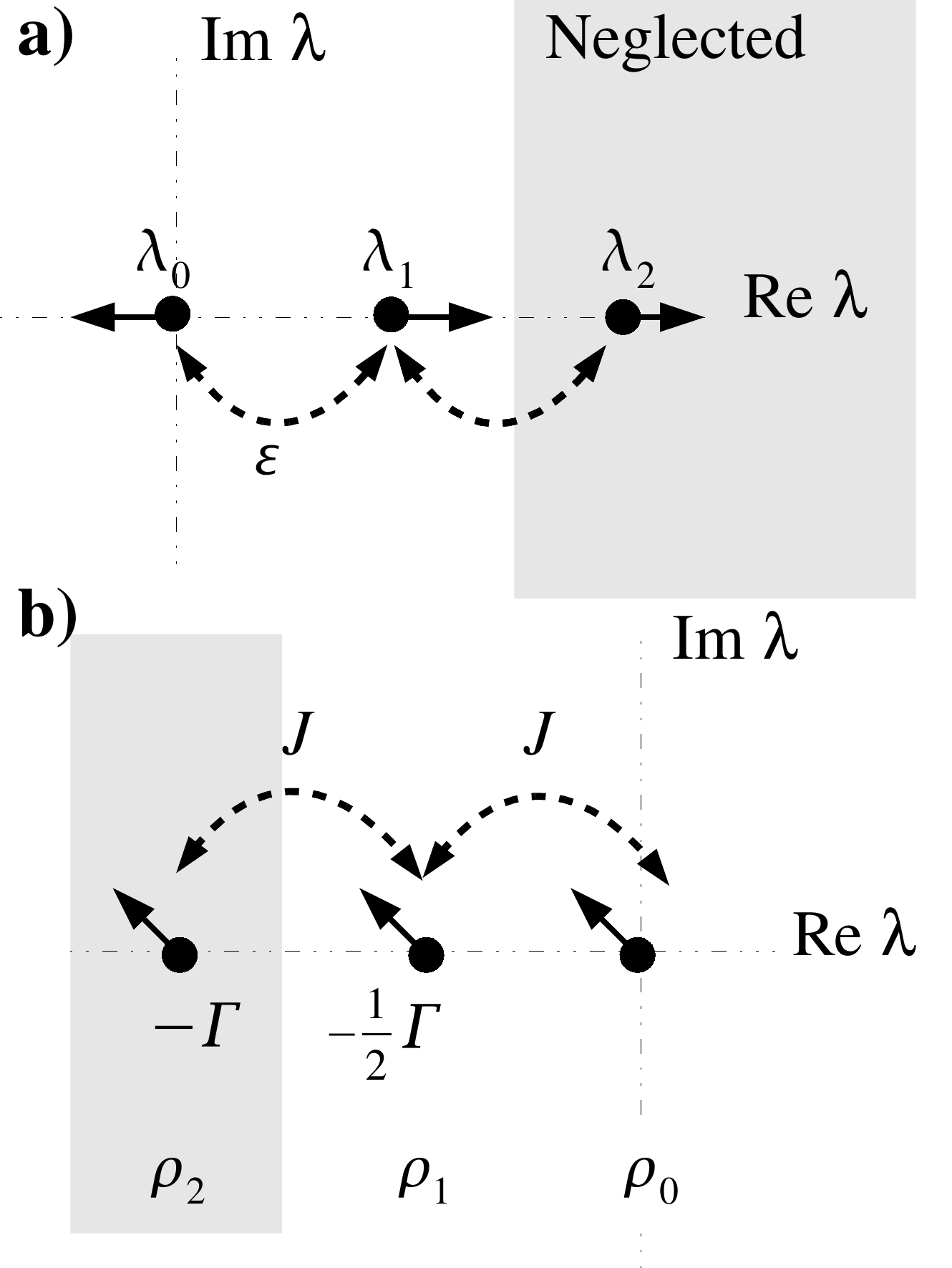}
\caption{\label{fig:spectrum} (a) In ordinary perturbation theory of
  Hermitian operators we find eigenspaces with eigenvalues which are
  well separated along the real axis. A weak Hermitian coupling,
  smaller than the energy separation between spaces $\varepsilon \ll
  |\lambda_i - \lambda_j|,$ only causes small shifts in these energy
  levels. (b) Our Liouville operator $\Lg$ is not Hermitian but has
  well separated eigenvalues in the space of density matrices (We have
  set $U_r=0$ for simplicity). The real part of these eigenvalues now
  represents the decay rate of such matrices. The hopping of particles
  $J$ acts as a weak perturbation $J \ll \hbar \Gamma,$ that slightly
  changes both the real and the imaginary part of the original
  eigenvalues. In particular, the states with one particle per site,
  $\rho_0,$ now acquire contributions of ${\cal O}(J/\Gamma)^1$ and
  ${\cal O}(J/\Gamma)^{2,3,\ldots}$ by coupling directly, $\rho_1,$ or
  indirectly, $\rho_{2,3,\ldots},$ to faster decaying subspaces. Both
  in (a) and (b), for second order approximations of the unperturbed
  eigenspaces, $\lambda_0,$ we may neglect all spaces with an indirect
  coupling.}
\end{figure}

Both $\Lv$ and $\Lg$ are linear operators that act on an appropriate
space of matrices.  Even though within this space the superoperators
are not Hermitian, the dominant term $\Lg$ has infinitely many
eigenstates forming a discrete spectrum of well separated points
beginning at $\lambda_0=0$ and spanning through the left half of the
complex plane [Fig.~\ref{fig:spectrum}b]. Each of these points
represents a family of matrices that, in the absence of hopping term,
decay at a rate given by the real part of the eigenvalue,
$\mathrm{Re}(\lambda_i) \leq 0.$ In the limit of strong dissipation,
the hopping $\Lv$ will couple these eigenspaces very weakly, with an
amplitude $J/\hbar$ much smaller than the typical eigenvalue
separation, $\Gamma.$

Precisely in this limit of weak hopping we will be able to write
analytic expansions of the eigenstates, eigenvalues and evolution
equations for the perturbed superoperator $(\Lv + \Lg).$ The ideas is
to use Kato's resolvent method \cite{kato:80} with an expansion of the
Liouvillian
\begin{equation}
  \Lg = \sum_i \lambda_i \Proj_i,
\end{equation}
which uses a complete set of pseudo-projector operators
\begin{equation}
  \Proj_i \Proj_j = \delta_{ij} \Proj_i,\quad
  \sum_i \Proj_i= 1.
  \label{completeness}
\end{equation}
These operators, built from the right and left eigenvectors of $\Lg,$
are non-Hermitian but this fact poses no difficulties in generalizing
the usual perturbation theory.

In particular, the space of density matrices that had zero decay rate
for $J=0$ will transform into the perturbed eigenstates and eigenvalues
\begin{equation}
  (\Lv + \Lg) \tilde \rho_0(J) = \tilde\lambda_0(J)\tilde\rho_0(J).
\end{equation}
The decay rate will no longer be zero but remain small
\begin{equation}
  \tilde\lambda_0(J) = J\sum_{n>1}
  c_n\left(\frac{J}{\hbar \Gamma}\right)^n \sim {\cal O}(J^2/\hbar\Gamma),
  \label{eigenvalues}
\end{equation}
and these states will acquire a small admixture of the matrices
belonging to the unperturbed eigenspaces that had larger decay rates
\begin{equation}
  \tilde\rho_0(J) =
  \rho_0 + \sum_{n>1} \left(\frac{J}{\hbar \Gamma}\right)^n\rho_n.
  \label{eigenstate}
\end{equation}
Notice the weight of the different contributions $\rho_n$ depends on
how many applications of $\Lv$ we have to perform to connect $\rho_0$
to these other eigenspaces. Furthermore, we understand that it is the
coupling of $\rho_0$ to the decaying states through $\Lv$ what makes
the initially stable space lose particles. A similar reasoning can be
applied to the rapidly decaying eigenspaces,
$\tilde\rho_{1,2,\ldots}(J),$ which will be modified by the coupling
$\Lv.$ However in this case the decay rates will remain of order
$\Gamma$ with small corrections from the hopping of particles.

We therefore conclude that if we prepare our particles in any initial
state, after a short transient $t\sim 1/\Gamma,$ most of the state
will be captured by the eigenspace with the lowest decay rate
(\ref{eigenvalues}). Following the structure given in
Eq.~(\ref{eigenstate}) we will decompose our evolved state in a
series of contributions from the unperturbed eigenspaces
\begin{equation}
  \rho(t) = \rho_0(t) + \rho_1(t) + \rho_2(t) + \ldots,
  \label{evolved}
\end{equation}
where, according to Eq.~(\ref{eigenstate}), the high order
contributions have a vanishingly small size
\begin{equation}
|\rho_n(t)|\sim (J/\hbar\Gamma)^n|\rho_0(t)|,\quad t\gg J.
\end{equation}
We are therefore allowed to perform a self-consistent approximation
which consists on, first, writing the evolution equations for each
subspace $\rho_n(t),$ then impose that all contributions $\rho_{n\geq
  2}\simeq 0$ and finally integrate the remaining equations until we
reach an effective model for the lowest order term.

\subsection{Local Projections}
\label{sec:projectors}

As sketched before, our study of the losses in the lattice begins by
finding out the eigenspaces of the unperturbed superoperator, $\Lg.$
This task is greatly simplified by the fact that
$\Lg=\sum_k{\Lloc}_{,k}$ is a sum of commuting local superoperators
${\Lloc}.$ We can thus focus on diagonalizing one of these
superoperators on a single lattice site.

We will now introduce some notation. Since we are interested in
density matrices as elements of a Hilbert space on which the
superoperators act, we will introduce a basis of vectors of this
space. For a single site we will define the Fock states
\begin{equation}
  \kket{n,m} = \frac{1}{\sqrt{n!m!}} a^{\dagger n}\ket{0}\bra{0}a^{m},
\end{equation}
and introduce the Hermitian conjugate $\bbra{n',m'}$ given by
$\bbrakket{n',m'}{n,m}=\delta_{nn'}\delta_{mm'}$. In this basis
$\Lloc$ becomes a bidiagonal, nonsymmetric operator
\begin{eqnarray}
  \lefteqn{
  \Lloc \kket{n,m} = -\frac{i}{2\hbar}U_r (\xi_n-\xi_m) \kket{n,m} 
  }\nonumber \\
  && - \frac \Gamma 4 (\xi_n+\xi_m) \kket{n,m} + \frac \Gamma 2 \sqrt{\xi_n \xi_m} \kket{n-2,m-2},
\end{eqnarray}
where $\xi_x=x(x-1)$. The kernel of this operator is given by states
with 0 or one particle per site, so that we can write the space of
non-decaying density matrices as in Eq.~(\ref{stable}).

In addition to the right eigenvectors,
$(\Lloc-\lambda_n)\kket{v_n}=0,$ we will also search the left
eigenvectors, $\bbra{w_n}(\Lloc-\lambda_n)=0,$ which have the property
$\bbrakket{w_n}{v_m}=\delta_{nm}.$ With these families of operators we
will construct a set of pseudo-projector operators of the form
\begin{equation}
  \Ploc_n = \kket{v_n}\bbra{w_n}.
\end{equation}

We face one problem, though, which is that $\Lloc$ acts on an
infinite-dimensional space, with occupation numbers that can be
arbitrarily large. We argue that for our purposes it suffices to
truncate the Hilbert space to occupations $n,m\leq 2.$ The reason is
that our initial states will all belong to the space $\rho_0$ given
above and, since we will neglect contributions from third and
subsequent order couplings [i.~e. $\rho_{2,3,\ldots}=0$ in
Eq.~(\ref{evolved})], we will obtain at most a double occupation per
site. The self-consistency of our approximation will be evident at the
end.

Using the previous truncation, $\Lloc$ becomes a $9\times9$
block-diagonal and banded matrix with projections onto eigenspaces
\begin{subequations}
\begin{eqnarray}
  \label{P-0-loc}
  \Ploc_0 &=& \kket{0,0}\bbra{2,2} + \sum_{b,b'=0,1} \kket{b,b'}\bbra{b,b'}\\
  \Ploc_{1a} &=& \sum_{b=0,1} \kket{2,b}\bbra{2,b} \\
  \Ploc_{1b} &=& \sum_{b=0,1} \kket{b,2}\bbra{b,2} \\
  \Ploc_2 &=& \kket{2,2}\bbra{2,2}-\kket{0,0}\bbra{2,2}
\end{eqnarray}
\end{subequations}
and corresponding eigenvalues
\begin{subequations}
\begin{eqnarray}
  \lambda_0 &=& 0 \\
  \label{lambda-1-2}
  \lambda_{1a} &=&   -\frac{i U}\hbar= -\frac\Gamma2-\frac{iU_r}\hbar \\
  \lambda_{1b} &=& \lambda_{1a}^*\\
  \lambda_2 &=& -\Gamma .
\end{eqnarray}
\end{subequations}
$\Ploc_{1a}$ and $\Ploc_{1b}$ are connected to $\Ploc_0$ by a single
application of the hopping term, $\Lv,$ and $\Ploc_2$ by two
applications of $\Lv.$ Note also that $\Ploc_0$ contains
not only a projector onto single or zero occupancy, but also a term
that destroys two particles at a site, emptying it. This will be
essential later on.

\subsection{Adiabatic elimination}
\label{sec:calc-effective}

As explained in Section~\ref{sec:effective}, we will consider that our
states can be described by a density matrix with contributions coming
from the eigenspace with the lowest decay rate, $\rho_0,$ and the two
eigenspaces connected to it, $\rho_{1a,1b}.$ These contributions are
respectively obtained by applying the following pseudo-projectors onto
$\rho(t)$
\begin{subequations}
\begin{eqnarray}
  \Proj_0 &=& \Ploc_0 \otimes \cdots \otimes \Ploc_0,\\
  \Proj_{1a} &=& \sum_{m=0}^{L-1} (\Ploc_0)^{\otimes m}\otimes \Ploc_{1a}
  \otimes (\Ploc_0)^{\otimes L-m-1},\\
  \Proj_{1b} &=& \sum_{m=0}^{L-1} (\Ploc_0)^{\otimes m}\otimes \Ploc_{1b}
  \otimes (\Ploc_0)^{\otimes L-m-1}.
\end{eqnarray}
\end{subequations}
Connecting to the previous notation of projectors onto states with zero
and one particles per site, it will be useful to realize that the
zeroth order projector can be written as follows
\begin{equation}
  \label{P0}
  \Proj_0\rho =
  Q_0\rho Q_0 + \frac{1}{2}\sum_k a^2_k Q_1 \rho Q_1 a^{\dagger 2}_k.
\end{equation}

Since we will neglect higher order couplings, we can use
Eqs.~(\ref{master-eq-V-Lint}) and (\ref{completeness}) to write
evolution equations for the density matrices in the form
\begin{subequations}
\begin{eqnarray}
  \label{splitting-0}
  \frac{d\rho_0}{dt} &=& \Lv_{00} \rho_0 + \sum_c \Lv_{0c} \rho_c \\
  \label{splitting-s}
  \frac{d\rho_c}{dt} &=& \lambda_{c} \rho_{c} + \Lv_{c0} \rho_0.
\end{eqnarray}
\end{subequations}
We have abbreviated $\Lv_{ij}= \Proj_i \Lv \Proj_j$ and introduced the
notation that the index $c$ runs through $\{1a,1b\}.$ The terms
$\rho_{c}$ can be integrated out of the model using the fact that our
states are initially prepared in the slow decaying manifold
$\rho_c(0)=0.$ Formal integration of Eq.~(\ref{splitting-s}) yields
\begin{eqnarray}
  \rho_c(t) &=& e^{\lambda_c t} \int_0^t d\tau e^{-\lambda_c \tau} \Lv_{c0} \rho_0(\tau),
\end{eqnarray}
which after integration by parts becomes
\begin{eqnarray}
  \rho_c(t) &=& - \frac 1{\lambda_c} \Lv_{c0} \left[ \rho_0(t) - e^{\lambda_c t} \rho_0(0)\right] \nonumber \\
  && + \frac{e^{\lambda_c t}}{\lambda_c} \int_0^t d\tau e^{-\lambda_c \tau} \Lv_{c0} \frac{d\rho_0}{dt}(\tau) .
\end{eqnarray}
We neglect the remaining integral, because it is of higher order in
$J/\hbar\Gamma$ than the previous term, which is evident from the fact
that $d\rho_0/dt \propto J$ in Eq.\ (\ref{splitting-0}). Insertion of
$\rho_c(t)$ into Eq.\ (\ref{splitting-0}) yields
\begin{eqnarray}
  \label{final-effective-eq}
  \frac{d\rho_0}{dt} &=& \left( \Lv_{00} - \sum_{c} \frac 1{\lambda_c} \Lv_{0c} \Lv_{c0} \right) \rho_0(t) \nonumber \\
  && + \sum_{c} \frac 1{\lambda_c} e^{\lambda_c t} \Lv_{0c} \Lv_{c0} \rho_0(0) .
\end{eqnarray}
The first line of this equation represents our effective model
(\ref{effective-model}) and will be discussed in the following
section. The second line is a transient that decays at a rate $\propto
\Gamma.$ Therefore we obtained that the system converges to the
slow-decaying eigenspace in a time $t\sim 1/\Gamma,$ much shorter than
the typical time scale $\hbar/J$ at which the effective model
operates.

\subsection{Hard-core bosons}
\label{sec:calc-tonks}

Let us analyze the lowest order contribution to our effective model
(\ref{final-effective-eq}), given by $\Lc_1=\Lv_{00}$. Using the
expression in Eq.~(\ref{P0}), we obtain
\begin{eqnarray}
  \Lc_1 \rho_0 &=& \Proj_0 \Lv \Proj_0 \rho_0 \nonumber \\
  &=& Q_0 \frac{-i}\hbar[H_{J}, Q_0 \rho_0 Q_0] Q_0 \nonumber \\
  &=& -\frac{i}{\hbar}[Q_0 H_{J} Q_0, \rho_0].
\end{eqnarray}
In other words, this Liouville operator is equivalent to a Hamiltonian
in which we have projected out all states with double occupation. This
is the hard-core bosons or Tonks-Girardeau gas model presented
in Sect.~\ref{sec:tonks}.

\subsection{Second order losses}
\label{sec:calc-losses}

We are now going to consider the second order Liouvillian $\Lc_2$ from
Eq.~(\ref{effective-model})
\begin{equation}
  \Lc_2 = \sum_{c\in\{1a,1b\}} \frac{-1}{\lambda_c}
  \Proj_0 \Lv \Proj_c \Lv \Proj_0.
  \label{L2}
\end{equation}
We can expand this expression
\begin{eqnarray}
  \Lc_2\rho_0 &=& \frac{-i^2}{\lambda_{1a}\hbar^2}\Proj_0 [H_J,
  Q_1[H_J, \rho_0]Q_0] + \\
  &+& \frac{-i^2}{\lambda_{1b}\hbar^2}\Proj_0 [H_J,
  Q_0[H_J, \rho_0]Q_1].\nonumber
\end{eqnarray}
Using the property $\rho_0=Q_0\rho_0Q_0,$ one realizes that the only
relevant terms are $\Lc_2\rho_0 = \Proj_0 {\cal A}\rho_0/\hbar^2$ with
\begin{eqnarray}
  {\cal A}\rho_0 &=& \frac{1}{\lambda_{1a}}\left(
    H_JQ_1H_JQ_0\rho_0 -
    Q_1H_JQ_0\rho_0Q_0H_J\right) \nonumber\\
  &+&\frac{1}{\lambda_{1b}}\left(
    Q_0\rho_0Q_0H_JQ_1H_J - H_JQ_0\rho_0Q_0H_JQ_1
    \right).
\end{eqnarray}
We now consider the final projection with $\Proj_0.$ Following
Eq.~(\ref{P0}), this pseudoprojector contains two operations: the
first one keeps terms proportional to $Q_0H_JQ_1H_JQ_0,$ while the
second one acts on the terms that create a doubly occupied site on
each side of the density matrix, that is $Q_1H_JQ_0\rho_0Q_0H_JQ_1.$
Introducing $T=Q_1H_JQ_0/(-J),$
\begin{eqnarray}
  \Lc_2\rho_0 &=& \frac{J^2}{\hbar^2} \left(
   \frac{1}{\lambda_{1a}} T^\dagger T\rho_0 +
   \frac{1}{\lambda_{1a}^*} \rho_0 T^\dagger T
    \right) \\
  &-& \frac{2J^2}{\hbar^2}\left(\mathrm{Re}\frac{1}{\lambda_{1a}}\right)
  \frac{1}{2}\sum_k a^2_kT\rho_0 T^\dagger a^{\dagger 2}_k.\nonumber
\end{eqnarray}
It is now time to rewrite everything in terms of hard core boson
operators. We notice the following equivalence
\begin{equation}
  T = Q_1 \sum_{\langle k,l\rangle} a_k^\dagger a_l Q_0
  = \sum_{k} a^{\dagger 2}_k C_k.
\end{equation}
This arises from the fact that $Q_1$ projects onto a state with a single
pair. Therefore, the tunneling term only contributes with processes that
take two neighboring particles ($C_k$) and create a pair in one of the
sites. Notice that $C_k$ already enforces the projection $Q_0.$ Using
this notation we can simplify our expressions even further
\begin{eqnarray}
  a^2_k T &=& 2 C_k,\\
  T^\dagger T &=& 2 \sum_k C_k^\dagger C_k,
\end{eqnarray}
thus arriving to the final model
\begin{eqnarray}
  \Lc_2\rho_0 &=&
  \frac{2J^2}{\hbar^2}\sum_k \left(
  \frac{1}{\lambda_{1a}} C_k^\dagger C_k\rho_0 +
  \frac{1}{\lambda_{1a}^*} \rho_0 C_k^\dagger C_k
  \right)\nonumber\\
  &-& \frac{2J^2}{\hbar^2}2\mathrm{Re}\frac{1}{\lambda_{1a}}
  \sum_{k} C_k \rho_0 C^\dagger_k,
\end{eqnarray}
which is studied in Sect.~\ref{sec:losses}.

\section{Conclusion}
\label{sec:conclusions}

We have shown analytically and confirmed numerically that strong,
inelastic interactions can induce a Tonks gas dynamics for a cloud of
molecules trapped in an optical lattice. The particles act like
hard-core bosons, with dissipation playing the role of a strong
repulsion. This effective model is completed with a reduced loss rate,
$\gamma_{\rm eff} \propto J^2/\Gamma$ which is much slower than both
the tunneling amplitude, $J/\hbar,$ and the original loss rate,
$\Gamma.$

Even with the small losses, the state of the system can at all times
be described as an incoherent mixture of strongly correlated Tonks
gases with different total particle number. In this respect, being
based on the idea of using dissipation to create strong correlations,
our paper connects to recent works which suggest using dissipation to
engineer states and phase transitions \cite{verstraete08,diehl08}.

We acknowledge financial support of the German Excellence Initiative
via the program Nanosystems Initiative Munich and of the Deutsche
Forschungsgemeinschaft via SFB 631. J.~J.~G.-R. acknowledges financial
support from the Ramon y Cajal Program of the M.~E.~C. and the
projects FIS2006-04885 and CAM-UCM/910758.


\begin{thebibliography}{15}
\expandafter\ifx\csname natexlab\endcsname\relax\def\natexlab#1{#1}\fi
\expandafter\ifx\csname bibnamefont\endcsname\relax
  \def\bibnamefont#1{#1}\fi
\expandafter\ifx\csname bibfnamefont\endcsname\relax
  \def\bibfnamefont#1{#1}\fi
\expandafter\ifx\csname citenamefont\endcsname\relax
  \def\citenamefont#1{#1}\fi
\expandafter\ifx\csname url\endcsname\relax
  \def\url#1{\texttt{#1}}\fi
\expandafter\ifx\csname urlprefix\endcsname\relax\def\urlprefix{URL }\fi
\providecommand{\bibinfo}[2]{#2}
\providecommand{\eprint}[2][]{\url{#2}}

\bibitem[{\citenamefont{Girardeau}(1960)}]{girardeau:60}
\bibinfo{author}{\bibfnamefont{M.}~\bibnamefont{Girardeau}},
  \bibinfo{journal}{J. Math. Phys.} \textbf{\bibinfo{volume}{1}},
  \bibinfo{pages}{516} (\bibinfo{year}{1960}).

\bibitem[{\citenamefont{Lieb and Liniger}(1963)}]{lieb:63}
\bibinfo{author}{\bibfnamefont{E.~H.} \bibnamefont{Lieb}} \bibnamefont{and}
  \bibinfo{author}{\bibfnamefont{W.}~\bibnamefont{Liniger}},
  \bibinfo{journal}{Phys. Rev.} \textbf{\bibinfo{volume}{130}},
  \bibinfo{pages}{1605} (\bibinfo{year}{1963}).

\bibitem[{\citenamefont{{Paredes} et~al.}(2004)\citenamefont{{Paredes},
  {Widera}, {Murg}, {Mandel}, {F{\"o}lling}, {Cirac}, {Shlyapnikov},
  {H{\"a}nsch}, and {Bloch}}}]{paredes:04}
\bibinfo{author}{\bibfnamefont{B.}~\bibnamefont{{Paredes}}},
  \bibinfo{author}{\bibfnamefont{A.}~\bibnamefont{{Widera}}},
  \bibinfo{author}{\bibfnamefont{V.}~\bibnamefont{{Murg}}},
  \bibinfo{author}{\bibfnamefont{O.}~\bibnamefont{{Mandel}}},
  \bibinfo{author}{\bibfnamefont{S.}~\bibnamefont{{F{\"o}lling}}},
  \bibinfo{author}{\bibfnamefont{I.}~\bibnamefont{{Cirac}}},
  \bibinfo{author}{\bibfnamefont{G.~V.} \bibnamefont{{Shlyapnikov}}},
  \bibinfo{author}{\bibfnamefont{T.~W.} \bibnamefont{{H{\"a}nsch}}},
  \bibnamefont{and} \bibinfo{author}{\bibfnamefont{I.}~\bibnamefont{{Bloch}}},
  \bibinfo{journal}{\nat} \textbf{\bibinfo{volume}{429}}, \bibinfo{pages}{277}
  (\bibinfo{year}{2004}).

\bibitem[{\citenamefont{Kinoshita et~al.}(2004)\citenamefont{Kinoshita, Wenger,
  and Weiss}}]{kinoshita:04}
\bibinfo{author}{\bibfnamefont{T.}~\bibnamefont{Kinoshita}},
  \bibinfo{author}{\bibfnamefont{T.}~\bibnamefont{Wenger}}, \bibnamefont{and}
  \bibinfo{author}{\bibfnamefont{D.~S.} \bibnamefont{Weiss}},
  \bibinfo{journal}{Science} \textbf{\bibinfo{volume}{305}},
  \bibinfo{pages}{1125} (\bibinfo{year}{2004}).

\bibitem[{\citenamefont{Kinoshita et~al.}(2005)\citenamefont{Kinoshita, Wenger,
  and Weiss}}]{kinoshita:05}
\bibinfo{author}{\bibfnamefont{T.}~\bibnamefont{Kinoshita}},
  \bibinfo{author}{\bibfnamefont{T.}~\bibnamefont{Wenger}}, \bibnamefont{and}
  \bibinfo{author}{\bibfnamefont{D.~S.} \bibnamefont{Weiss}},
  \bibinfo{journal}{Physical Review Letters} \textbf{\bibinfo{volume}{95}},
  \bibinfo{eid}{190406} (pages~\bibinfo{numpages}{4}) (\bibinfo{year}{2005}).

\bibitem[{\citenamefont{Syassen et~al.}(2008)\citenamefont{Syassen, Bauer,
  Lettner, Volz, Dietze, Garcia-Ripoll, Cirac, Rempe, and Durr}}]{syassen:08}
\bibinfo{author}{\bibfnamefont{N.}~\bibnamefont{Syassen}},
  \bibinfo{author}{\bibfnamefont{D.~M.} \bibnamefont{Bauer}},
  \bibinfo{author}{\bibfnamefont{M.}~\bibnamefont{Lettner}},
  \bibinfo{author}{\bibfnamefont{T.}~\bibnamefont{Volz}},
  \bibinfo{author}{\bibfnamefont{D.}~\bibnamefont{Dietze}},
  \bibinfo{author}{\bibfnamefont{J.~J.} \bibnamefont{Garcia-Ripoll}},
  \bibinfo{author}{\bibfnamefont{J.~I.} \bibnamefont{Cirac}},
  \bibinfo{author}{\bibfnamefont{G.}~\bibnamefont{Rempe}}, \bibnamefont{and}
  \bibinfo{author}{\bibfnamefont{S.}~\bibnamefont{Durr}},
  \bibinfo{journal}{Science} \textbf{\bibinfo{volume}{320}},
  \bibinfo{pages}{1329} (\bibinfo{year}{2008}).

\bibitem[{\citenamefont{Verstraete et~al.}(2004)\citenamefont{Verstraete,
  Garcia-Ripoll, and Cirac}}]{verstraete:04}
\bibinfo{author}{\bibfnamefont{F.}~\bibnamefont{Verstraete}},
  \bibinfo{author}{\bibfnamefont{J.~J.} \bibnamefont{Garcia-Ripoll}},
  \bibnamefont{and} \bibinfo{author}{\bibfnamefont{J.~I.} \bibnamefont{Cirac}},
  \bibinfo{journal}{Physical Review Letters} \textbf{\bibinfo{volume}{93}},
  \bibinfo{eid}{207204} (pages~\bibinfo{numpages}{4}) (\bibinfo{year}{2004}).

\bibitem[{\citenamefont{et~al}()}]{duerr:prep}
\bibinfo{author}{\bibfnamefont{D.~S.} \bibnamefont{et~al}}, \bibinfo{note}{in
  preparation}.

\bibitem[{\citenamefont{{Jaksch} et~al.}(1998)\citenamefont{{Jaksch}, {Bruder},
  {Cirac}, {Gardiner}, and {Zoller}}}]{jaksch:98}
\bibinfo{author}{\bibfnamefont{D.}~\bibnamefont{{Jaksch}}},
  \bibinfo{author}{\bibfnamefont{C.}~\bibnamefont{{Bruder}}},
  \bibinfo{author}{\bibfnamefont{J.~I.} \bibnamefont{{Cirac}}},
  \bibinfo{author}{\bibfnamefont{C.~W.} \bibnamefont{{Gardiner}}},
  \bibnamefont{and} \bibinfo{author}{\bibfnamefont{P.}~\bibnamefont{{Zoller}}},
  \bibinfo{journal}{Phys. Rev. Lett.} \textbf{\bibinfo{volume}{81}},
  \bibinfo{pages}{3108} (\bibinfo{year}{1998}).

\bibitem[{\citenamefont{Freericks and Monien}(1996)}]{freericks:96}
\bibinfo{author}{\bibfnamefont{J.~K.} \bibnamefont{Freericks}}
  \bibnamefont{and} \bibinfo{author}{\bibfnamefont{H.}~\bibnamefont{Monien}},
  \bibinfo{journal}{Phys. Rev. B} \textbf{\bibinfo{volume}{53}},
  \bibinfo{pages}{2691} (\bibinfo{year}{1996}).

\bibitem[{\citenamefont{{Volz} et~al.}(2006)\citenamefont{{Volz}, {Syassen},
  {Bauer}, {Hansis}, {D{\"u}rr}, and {Rempe}}}]{volz:06}
\bibinfo{author}{\bibfnamefont{T.}~\bibnamefont{{Volz}}},
  \bibinfo{author}{\bibfnamefont{N.}~\bibnamefont{{Syassen}}},
  \bibinfo{author}{\bibfnamefont{D.~M.} \bibnamefont{{Bauer}}},
  \bibinfo{author}{\bibfnamefont{E.}~\bibnamefont{{Hansis}}},
  \bibinfo{author}{\bibfnamefont{S.}~\bibnamefont{{D{\"u}rr}}},
  \bibnamefont{and} \bibinfo{author}{\bibfnamefont{G.}~\bibnamefont{{Rempe}}},
  \bibinfo{journal}{Nature Physics} \textbf{\bibinfo{volume}{2}},
  \bibinfo{pages}{692} (\bibinfo{year}{2006}).

\bibitem[{\citenamefont{Cazalilla}(2003)}]{cazalilla03}
\bibinfo{author}{\bibfnamefont{M.~A.} \bibnamefont{Cazalilla}},
  \bibinfo{journal}{Phys. Rev. A} \textbf{\bibinfo{volume}{67}},
  \bibinfo{pages}{053606} (\bibinfo{year}{2003}).

\bibitem[{\citenamefont{Kato}(1995)}]{kato:80}
\bibinfo{author}{\bibfnamefont{T.}~\bibnamefont{Kato}},
  \emph{\bibinfo{title}{Perturbation Theory for Linear Operators}}
  (\bibinfo{publisher}{Springer-Verlag}, \bibinfo{address}{Berlin},
  \bibinfo{year}{1995}), chap.~\bibinfo{chapter}{II}.

\bibitem[{\citenamefont{{Verstraete} et~al.}(2008)\citenamefont{{Verstraete},
  {Wolf}, and {Cirac}}}]{verstraete08}
\bibinfo{author}{\bibfnamefont{F.}~\bibnamefont{{Verstraete}}},
  \bibinfo{author}{\bibfnamefont{M.~M.} \bibnamefont{{Wolf}}},
  \bibnamefont{and} \bibinfo{author}{\bibfnamefont{J.~I.}
  \bibnamefont{{Cirac}}}, \bibinfo{journal}{ArXiv e-prints}
  \textbf{\bibinfo{volume}{803}} (\bibinfo{year}{2008}), \eprint{0803.1447}.

\bibitem[{\citenamefont{{Diehl} et~al.}(2008)\citenamefont{{Diehl}, {Micheli},
  {Kantian}, {Kraus}, {B{\"u}chler}, and {Zoller}}}]{diehl08}
\bibinfo{author}{\bibfnamefont{S.}~\bibnamefont{{Diehl}}},
  \bibinfo{author}{\bibfnamefont{A.}~\bibnamefont{{Micheli}}},
  \bibinfo{author}{\bibfnamefont{A.}~\bibnamefont{{Kantian}}},
  \bibinfo{author}{\bibfnamefont{B.}~\bibnamefont{{Kraus}}},
  \bibinfo{author}{\bibfnamefont{H.~P.} \bibnamefont{{B{\"u}chler}}},
  \bibnamefont{and} \bibinfo{author}{\bibfnamefont{P.}~\bibnamefont{{Zoller}}},
  \bibinfo{journal}{ArXiv e-prints} \textbf{\bibinfo{volume}{803}}
  (\bibinfo{year}{2008}), \eprint{0803.1482}.

\end{thebibliography}
\end{document}